\documentclass[11pt,a4paper]{article}
\pdfoutput=1

\usepackage{amsmath}
\usepackage{amsfonts}
\usepackage{amsthm}
\usepackage{amssymb}
\usepackage{amscd}
\usepackage[english]{babel}
\usepackage{graphicx}
\usepackage{psfrag}
\usepackage{epsfig}
\usepackage{rotating}
\usepackage{times}
\usepackage{color}
\usepackage{ulem}
\usepackage{longtable}

\theoremstyle{plain}

\newtheorem{remark}{Remark}[section]

\newcommand{\boxend}{\flushright{$\Box$}}

\begin{document}

\title{Discrepancies between observational data and theoretical forecast in  single field slow roll inflation}


\author{Jaume Amor\'os$^{a,}$\footnote{E-mail: jaume.amoros@upc.edu} and Jaume de Haro$^{a,}$\footnote{E-mail: jaime.haro@upc.edu}}

\maketitle
 
\begin{center}
{\small
$^a$Departament de Matem\`atica Aplicada I, Universitat
Polit\`ecnica de Catalunya \\ Diagonal 647, 08028 Barcelona, Spain \\
}
\end{center}

%
%





\thispagestyle{empty}

\begin{abstract}
The PLANCK collaboration has determined,
or greatly constrained, values 
for the spectral parameters of the CMB radiation, 
 namely the spectral index $n_s$, its
running $\alpha_s$, the running of the running $\beta_s$,
using a growing body of measurements of CMB anisotropies 
by the Planck satellite and other missions. 
These values do not
follow the hierarchy of sizes predicted by single field,
slow roll inflationary theory, and are thus difficult to
fit for such inflation models.

In this work we present first a study of 49 single field, 
slow roll inflationary potentials in which we assess the 
likelyhood of these models fitting the spectral parameters
to their currently most accurate determination
given by the PLANCK collaboration. 
We check numerically with a MATLAB program
the spectral parameters that each model can yield for a very broad,
comprehensive list of possible parameter and field values.
The comparison of spectral parameter values supported by the models
with their determinations by the PLANCK collaboration
leads to the conclusion that
the data provided by PLANCK2015 TT+lowP and PLANCK2015 TT,TE,EE+lowP
 taking into account the running of the running disfavours
40 of the 49 models with confidence level at least 92.8\%.

Next, we discuss the reliability of the current computations
of these spectral parameters. We identify a bias
in the method of determination of the spectral parameters 
by least residue parameter fitting (using MCMC or any other scheme)
currently
used to reconstruct the power spectrum of scalar perturbations.
This bias can explain the observed contradiction between theory and observations. Its removal is computationally costly, but necessary in order
to compare the forecasts of single field, slow roll theories
with observations.
\end{abstract}



\maketitle

\section{Introduction}

Recent astronomical projects, notably the ongoing Planck collaboration, 
are mapping with ever greater detail the anisotropies of the CMB radiation,
and determining the spectral parameters on which this radiation depends
with increasing accuracy. Such a trove of data allows the submission
of cosmological models to the test of comparing their forecasts 
with the observations.

The aim of this work is such a comparison. It concerns on one hand 
the power spectrum of the curvature fluctuation in comoving coordinates of the CMB
$$
{\mathcal P}_{\mathcal R}(k)= A_s \left( \frac{k}{k_*} \right)^{n_s-1+
\frac{1}{2} \alpha_s \ln(\frac{k}{k_*})+ \frac{1}{6} \beta_s \ln^2(\frac{k}{k_*})+ \dots} ,
$$
 which governs
the better discerned parts of the CMB and has been greatly constrained
by the Planck collaboration (Ade et al. 2013, 2015), and on
the other hand a
family of 49 of the best known single field, slow roll inflationary models
as compiled and systematized in (Martin, Ringeval \& Vennin 2013) (see its Table 1).

Every one of the discussed slow roll inflation models admits a range 
of possible values for the slow roll parameters.
Through the formulas recalled in Section \ref{s:formules}, these parameters determine the successive coefficients of the Taylor series expansion
of $\text{ln}\, {\mathcal P}_{\mathcal R}(k)$ at the chosen pivot scale $k_*$: 
the spectral index $n_s$, its running $\alpha_s$, the running of the running
$\beta_s$, \dots

These spectral parameters $n_s, \alpha_s, \beta_s$ follow an approximately
 Gaussian distribution, and the Planck collaboration has 
produced determinations of their expected values and deviations. Therefore,
it is a straightforward question for each single field, slow roll model
to find out the possible range of values that it supports for these 
parameters, and check at what distance they lie from their values
as determined by Planck.

The goal of our study is 
conceptually and practically simpler: to subject each model individually
to the comparison between its forecast of spectral parameter values
and the values actually found by (Ade et al. 2013, 2015), which
give the most precise determination up to date. On the other hand, 
let us remark that is not the aim of this work  to perform a Bayesian
comparison between the models, as in (Martin, Ringeval \& Trotta 2014), or to apply a Bayesian test
of relevance to the spectral parameters as in (Giannantonio \& Komatsu 2015).

The interest of such a simple analysis of model forecasts vs observations
can be inferred from the values that (Ade et al. 2013, 2015)
atribute to the running $\alpha_s$ and running of the running $\beta_s$.
According to single field, slow roll inflation $n_s-1$ is of order 1,
$\alpha_s$ of order 2, and $\beta_s$ of order 3 in the inflation parameters
$\epsilon,\eta$ (see Sect. \ref{s:formules}). 
Yet in all the determinations of the Planck team the running 
$\alpha_s$, or the running of the running $\beta_s$ have the same order 
of magnitude as $n_s-1$, namely $10^{-2}$. This is a hint of a strong disagreement
between theoretical forecasts and observations.

The causes for this discordance of orders of magnitude may lie 
in the lack of precission either of the theoretical models or of 
the observed, actually computed from CMB measurements, values
for the spectral parameters. This is the topic discussed in Section \ref{s:reliability}, where it is found that there is a subtle mathematical
cause of inaccuracy in the computation of the values of the spectral 
parameters in the highest-likelyhood fits currently used.

This is not a problem with the employed statistical techniques,
but with the mathematical meaning of the spectral parameters:
they are the coefficients of the Taylor expansion of the function
$\ln {\mathcal P}_{\mathcal R}(k)$ at a pivot scale $k_*$, thus they yield
the best possible approximation of this function with a polynomial 
of a specified degree {\it in a suitably small neighbourhood of $k_*$}. 
In contrast, any likelyhood-maximizing method seeks the polynomial
of the specified degree that best approximates $\ln {\mathcal P}_{\mathcal R}(k)$
{\it in a fixed interval $[k_*,k_{final}]$}. 
The regression polynomial of a specified degree 
for a function $g(k)$ in a fixed interval $[k_*,
k_{final}]$, and the likelyhood-maximizing polynomials that
can be characterized as regression polynomials with a suitable weighted
inner product, are different from the Taylor polynomial of $g(k)$ at
the point $k_*$. In Sect. \ref{s:reliability} we show, both with an
elementary example using linear least square fitting and with an 
actual computation of spectral parameters by residue minimization, 
 how the order
of magnitude discrepancy between the coefficients of the two polynomials
often mimics closely that between Planck's evaluations and the theoretical
determinations of the running $\alpha_s$ and the running of the running
$\beta_s$. While this interpretation flaw does not affect a Bayesian
discussion of whether higher order terms in $\ln {\mathcal P}_{\mathcal R}(k)$
are necessary for the description of the CMB power spectrum,
it has to be solved in order to obtain a meaningful comparison 
between theoretical models and observations.

\medskip

The units used in the paper are: $\hbar=c=8\pi G=1.$

\medskip

\section{ Slow-roll parameters} \label{s:formules}
In  slow roll inflation (see Basset, Tsujikawa \& Wands (2006) for a review of inflation) the commonly used  first order parameters are:
\begin{eqnarray}\label{x14}
\epsilon= -\frac{\dot{H}}{H^2}\cong \frac{1}{2}\left(\frac{V_{\varphi}}{V} \right)^2 \quad\mbox{and}\quad \eta=2\epsilon-\frac{\dot{\epsilon}}{2H\epsilon}
\cong\frac{V_{\varphi\varphi}}{V}.
\end{eqnarray}

At the first slow roll order,
the spectral index  of scalar perturbations and its running are given by
\begin{eqnarray}\label{x15}
 {n}_s-1=2\eta-6\epsilon\quad\mbox{and}\quad  {\alpha}_s=16\epsilon\eta-24\epsilon^2-2\xi,
\end{eqnarray}
where  the second order slow roll parameter
\begin{eqnarray}
\xi\equiv \left(2\epsilon-\frac{\dot\eta}{{ H}\eta}\right)
\eta\cong \frac{V_{\varphi}V_{\varphi\varphi\varphi}}{V^2},
\end{eqnarray}
has been introduced.

Moreover, in inflationary cosmology, the tensor/scalar ratio, namely $r$,  is related with the slow roll parameter $\epsilon$, via the following consistency relation
${r}=16\epsilon.$

The other important parameter that we will use in this work is the running of the running
$\beta_s=\frac{d \alpha_s}{d \ln k}$, given, in the slow roll approximation, by  (Huang 2006; Ade et al. 2013) 
\begin{eqnarray} \label{eq:betas}
 \beta_s=-192{\epsilon}^3+192{\epsilon}^2\eta-32\epsilon{\eta}^2-24\epsilon\xi+2\eta\xi+2\zeta,
\end{eqnarray}
where we have introduced the third order slow roll parameter
\begin{eqnarray} \label{eq:zeta}
 \zeta\equiv \left(4\epsilon-\eta-\frac{\dot{\xi}}{H\xi}\right)\xi\cong \frac{V^2_{\varphi}V_{\varphi\varphi\varphi\varphi}}{V^3}.
\end{eqnarray}

\medskip

\section{PLANCK2015 data: the running of the running}
The last PLANCK2015 data about the running $\alpha_s$ and its  running $\beta_s=
\frac{d \alpha_s}{d \ln k}$ are reproduced in Table 1.

\begin{table}[ht]
\label{t:planck2015betas}
\begin{tabular}{|l|c|c|c|}
\hline
Determination & $n_s$ & $\alpha_s$ & $\beta_s$ \\
\hline
PLANCK2015 TT+lowP & $0.9569\pm 0.0077$ & $0.011^{+0.014}_{-0.013}$ &
$0.029^{+0.015}_{-0.016}$ \\
\hline
PLANCK2015 TT,TE,EE+lowP & $0.9586\pm 0.0056$ & $0.009\pm 0.010$ & $0.025\pm 0.013$ \\
\hline
\end{tabular}
\caption{Determinations, or constrains in case of being nontrivial, of the spectral parameter values, with running of the running by PLANCK2015 ((19) of Ade et al. (2015))}
\end{table}

These results contradict single field slow roll inflation, because in that case, the running $\alpha_s$ is second order in the slow roll parameters, and its running is given by Eqs.
\eqref{eq:betas}, \eqref{eq:zeta},
which make $\beta_s$ a third order parameter, while the values determined by
PLANCK place $\beta_s$ in a higher order of magnitude than the running $\alpha_s$
itself. Moreover, disregarding the running of the running, the running is negative while taking into account it, the running becomes positive. This seems a signature of
the problem that suffers the method used to reconstruct the power spectrum of scalar perturbations from observational data:
the value of the coefficients in the Taylor series of the power spectrum logarithm function could suffer a bias. We will address this question later.

To show the improbability of the observed value of $\beta_s$
analytically for all the potentials that appear in  (Martin, Ringeval \& Vennin 2013)
is very involved due to the increasing complexity of the formulas \eqref{eq:betas}, \eqref{eq:zeta} for the new parameter $\beta_s$. 
However 
in the case of  LI (with $\alpha>0$),  SFI and  BI  (with $p$ and even number), HTI, ESI,  PLI and LFI,
a simple calculation shows that $\beta_s\leq 0$ which means that the deviation from the theoretical value of the running of
the running to its expected observational value  is  larger than $1.9\sigma$, using both, PLANCK2015 TT,TE,EE+lowP or PLANCK2015 TT+lowP data.
\medskip

The other potentials appearing in (Martin, Ringeval \& Vennin 2013) are analysed numerically.

\subsection{Numerical fitting of the parameters} 
\label{ss:algoritme}


Let us describe the numerical tests that the authors have applied to all single field inflationary models from the list of (Martin, Ringeval \& Vennin 2013). These tests have been 
built into a
MATLAB program that takes as input
a list of potentials $V(\varphi)$ and values for spectral parameters in the list $r,n_s,\alpha_s,\beta_s$,
and asseses the likelihood of each model in the list to fit the values of the running of the running $\beta_s$, assuming Gaussian distribution for all the spectral parameters.


For each cosmological model, a broad
range of possible values for the parameters on which it depends has been determined
following (Martin, Ringeval \& Vennin 2013).
A test list of values for each parameter has been selected, covering in a
dense, equispaced fashion finite intervals of possible values for the parameter,
and approaching with log-equispaced values every finite or infinite limit value
for the parameter.

The MATLAB software developed by the authors,
for each model $V(\varphi)$ and choice of value of the parameter(s) on which it depends,
takes an equispaced mesh of values in the range $[\varphi_0,\varphi_f]$ of possible values of the field in this model. This mesh is taken increasingly fine, currently up to
step $\Delta \varphi = 2 \cdot 10^{-4}$.

The subintervals in the range of field values for which the potential satisfies
$V(\varphi)>0$ are numerically determined over the selected mesh, and each
interval of positive values of the potential for the selected values
of the model parameters is considered as a {\em case}, which thus consists of:
\begin{itemize}
\item a candidate theory with a given potential $V(\varphi)$,
\item a specific choice of
parameter values for $V(\varphi)$,
\item and a range of values $[\bar \varphi_0,\bar \varphi_f]$ of the inflaton field $\varphi$
such that $V(\varphi)>0$ on them.
\end{itemize}

The numerical test for each case consists in meshing the interval of field
values with a uniform step (of size $\Delta \varphi = 2 \cdot 10^{-4}$ for the results reported
in this work), computing the spectral parameters $r, n_s, \alpha_s, \beta_s$
for each value of the field $\varphi$ in the mesh using the formulas
of Section \ref{s:formules} and symbolic derivation of the potential $V$
to produce the derivatives $V_\varphi, \dots, V_{\varphi \varphi \varphi \varphi}$,
 and then applying successive filtering criteria to determine
which values of the field $\varphi$ fulfill simultaneously all of them, thus allowing the
model in this particular case to fit the spectral measured data for which the model is tested.

\medskip

PLANCK2015  provides
the main values of the spectral paramenters, namely  $<n_s>,<\alpha_s>,<\beta_s>$, and their standard
deviations of the respective one-dimensional marginalized posterior distribution, namely $\sigma_{n_s},\sigma_{\alpha_s},\sigma_{\beta_s}$.

\medskip

The applied filters in the case of spectral parameters with running
of the running consist in looking for the values of the field $\varphi$ such that:
\begin{enumerate}
\item $\epsilon(\varphi) \le 1$,  $|\eta(\varphi)|\leq 1$, $|\xi(\varphi)|\leq 1$ and $|\zeta(\varphi)|\leq 1$.
\item The number of e-folds $N(\varphi)$ ranges between $50$ and $60$.
\item $|\beta_s(\varphi)-<\beta_s>| < 1.8\sigma_{\beta_s}$.
\end{enumerate}

Then, dealing with the one-dimensional marginalized posterior distribution of the running of the running, a model not passing this filter 
is ruled out with 92.8 \% C.L..


\begin{remark}
The testing software  looks for
values of the field $\varphi_e$ such that $\epsilon(\varphi_e) \cong 1$,
and using them as endpoints of the inflationary phase,
computes the number of e-folds of inflation for any choice of $\varphi$
in the case, by integrating numerically with a trapezoidal rule
\begin{eqnarray}
N(\varphi)=\left|\int_{\varphi_e}^{\varphi} \frac{V}{V_\varphi} d\varphi\right|.
\end{eqnarray}
\end{remark}

Finally, it is important to realize that in our analysis the  one-dimensional marginalized 92.8 \% C.L. interval for $\beta_s$ is
compared to the theoretical
predictions of  inflationary models,
and differs form the usual one where the  marginalized joint 95.5 \% C.L. region for ($n_s$, $r$) without a running spectral index 
is compared with the theoretical forecast.

\subsection{Numerical results} \label{ss:numresults}

Single-field inflaton models were exhaustively studied in (Martin, Ringeval \& Vennin 2013),
from which we take the list of models and parameters to be numerically tested.
Table \ref{t:llistamodels}, adapted from Table 1 of (Martin, Ringeval \& Vennin 2013),
presents each model's potential, the range of values of the parameters for which
it has been tested, and the range of values of the inflaton field over
which it has been tested.

{\tiny
\begin{longtable}{|c|c|c|c|}
\label{t:llistamodels}
  Name  & $V(\varphi)$ & Parameter values & Field values \\
  \hline \hline
  HI & $V_0\left(1-e^{-\sqrt{2/3}\varphi}\right)^2$ &  & [-40,40] \\
  \hline \hline
  RCHI & $V_0\left(1-2e^{-\sqrt{2/3}\varphi}+\frac{A_I}{16\pi^2}
 \frac{\varphi}{\sqrt{6}}\right)$ & $A_I$: [linspace(-100,100,120),linspace(-3,3,200)] & [-10,20] \\
  \hline \hline
  LFI & $V_0\left(\varphi\right)^p$ & $p$: linspace(0.5,20,60) &  \\
  \hline \hline
  MLFI & $V_0\varphi^2 \left[1 + \alpha \varphi^2 \right]$ & $\alpha$: [linspace(-10,100,61),linspace(-0.1,0.1,120)] &  \\
  \hline \hline
  RCMI & $V_0\left(\varphi\right)^2
\left[1-2\alpha\varphi^2\ln \left(\varphi\right)\right]$ & $\alpha$: [linspace(1e-4,1.5,30),10.$\wedge$linspace(-14,-5,20)] &  \\
  \hline \hline
  RCQI & $V_0\left(\varphi\right)^4
\left[1-\alpha \ln\left(\varphi \right)\right]$ & $\alpha$: [linspace(1e-2,10,120),10.$\wedge$linspace(-6,-2.5,10)] &  \\
  \hline \hline
  NI & $V_0\left[1+\cos\left(\frac{\varphi}{f}\right)\right]$ & $f$: 1 & $[10^{-5},\pi]$ \\
  \hline \hline
  ESI & $V_0\left(1-e^{-q\varphi}\right)$ & $q$: [linspace(0.1,10,120),linspace(1e-5,0.099,60)] &  \\
  \hline \hline
  PLI & $V_0e^{-\alpha \varphi}$ & $\alpha$: [linspace(0.1,10,120),10.$\wedge$linspace(-6,-1.5,15)] & [-40,40] \\
  \hline \hline
  KMII & $V_0\left(1-\alpha\varphi e^{-\varphi}\right)$ & $\alpha$: [linspace(1e-2,10,60),10.$\wedge$linspace(-6,-2.5,10)] &  \\
  \hline \hline
  HF1I & $V_0 \left(1+A_1 \varphi\right)^2\left[1-\frac{2}{3}
\left(\frac{A_1}{1+A_1\varphi}\right)^2\right]$ & $A_1$: linspace(1e-3,40,180) & [-40,40] \\
  \hline \hline
  CWI & $V_0\left[1 +
\alpha\left(\frac{\varphi}{Q}\right)^4 \ln
\left(\frac{\varphi}{Q}\right)\right]$ & $Q$: [10.$\wedge$linspace(-6,-2.5,12),linspace(1e-2,10,60)] &  \\
  \hline \hline
  LI & $V_0\left[1
+\alpha\ln \left(\varphi\right)\right]$ & $\alpha$: [linspace(-0.3,0.3,60),-10.$\wedge$linspace(-1,0,10),10.$\wedge$linspace(-1,0,10)] &  \\
  \hline \hline
  RpI & $V_0 e^{-2 \sqrt{2/3}\varphi} \left|e^{\sqrt{2/3}\varphi}
 - 1 \right|^{2p/(2p-1)}$  & $p$: linspace(0.25,10,60) &  \\
  \hline \hline
  DWI & $V_0\left[\left(\frac{\varphi}{\varphi_0}\right)^2-1\right]^2$ & $\varphi_0$: 1 & $[10^{-4},80]$ \\
  \hline \hline
  MHI & $V_0 \left[1-\text{sech} \left(\frac{\varphi}{\mu} \right) \right]$ & $\mu$: 10 & $[10^{-4},400]$ \\
  \hline \hline
  RGI & $V_0\frac{\left(\varphi\right)^2}{\alpha+\left(\varphi\right)^2}$ & $\alpha$: [linspace(1e-1,10,30),10.$\wedge$linspace(-6,-1.5,12)] &  \\
  \hline \hline
  MSSMI & $V_0\left[\left(\frac{\varphi}{\varphi_0}\right)^2-\frac{2}{3}
  \left(\frac{\varphi}{\varphi_0}\right)^6+\frac{1}{5}\left(
  \frac{\varphi}{\varphi_0}\right)^{10}\right]$ & $\varphi_0$: [1e-7,1e-3,1] & $[10^{-4},80]$ \\
  \hline \hline
  RIPI & $V_0 \left[ \left(\frac{\varphi}{\varphi_0}\right)^2 -
   \frac{4}{3} \left( \frac{\varphi}{\varphi_0} \right)^3 + \frac{1}{2}
   \left( \frac{\varphi}{\varphi_0} \right)^4 \right]$ & $\varphi_0$: [1e-7,1e-3,1] & $[10^{-4},80]$ \\
  \hline \hline
  AI & $V_0\left[1-\frac{2}{\pi}
    \arctan\left(\frac{\varphi}{\mu}\right)\right]$ & $\mu$: [1e-2,1] & [-40,40] \\
  \hline \hline
  CNAI & $V_0\left[3-\left(3+\alpha^2 \right) \tanh^2
    \left( \frac{\alpha}{\sqrt{2}} \varphi \right) \right]$ & $\alpha$: [linspace(1e-2,20,120),10.$\wedge$linspace(-6,-2.5,10)] &  \\
  \hline \hline
  CNBI & $V_0\left[\left(3-\alpha^2\right) \tan^2
    \left(\frac{\alpha}{\sqrt{2}}\varphi \right)-3\right]$ & $\alpha$: [linspace(1e-3,5,40),10.$\wedge$linspace(-7,-3.5,12)] &  \\
  \hline \hline
  OSTI & $-V_0\left(\frac{\varphi}{\varphi_0}\right)^2\ln\left[\left(\frac{\varphi}{\varphi_0}\right)^2\right]$ & $\varphi_0$: 1 & $[10^{-6},1]$ \\
    \hline \hline
  WRI & $V_0\ln^2 \left(\frac{\varphi}{\varphi_0}\right)$ & $\varphi_0$: 1 & $[10^{-4},10]$ \\
   \hline \hline
  SFI & $V_0 \left[1 -\left(\frac{\varphi}{\mu}\right)^{p}\right]$ & \begin{tabular}{l} $\mu$: 1 \\ $p$: linspace(0.5,10,20) \end{tabular} & $[10^{-4},1]$ \\
  \hline \hline
  II & $V_0\left(\varphi-\varphi_0\right)^{-\beta}
-V_0\frac{\beta ^2}{6}\left(\varphi-\varphi_0\right)^{-\beta-2}$ & \begin{tabular}{l} $\varphi_0$: 0 \\ $\beta$: [linspace(0.1,10,31),linspace(20,50,3)] \end{tabular} &  \\
  \hline \hline
  KMIII & $V_0\left[1-\alpha(\varphi)^{\frac43}\exp\left(-\beta(\varphi)^{\frac43}\right)\right]$ & \begin{tabular}{l} $\alpha$: 10.$\wedge$linspace(-3,12,46) \\ $\beta$: 10.$\wedge$linspace(-3,12,46) \end{tabular} & $[10^{-4},10]$ \\
  \hline \hline
  LMI & $V_0\left(\varphi\right)^{\alpha}\exp\left[-\beta (\varphi)^{\gamma}\right]$ & \begin{tabular}{l} $\beta$: [linspace(0.1,20,40),10.$\wedge$linspace(-4,-1.5,10)] \\ $\gamma$: [linspace(1e-3,2,30),10.$\wedge$linspace(0.5,2,4),10.$\wedge$linspace(-6,-4,3)] \end{tabular} &  \\
  \hline \hline
TWI & $V_0\left[1-A\left(\frac{\varphi}{\varphi_0}
    \right)^2e^{-\varphi/\varphi_0} \right]$ & \begin{tabular}{l} $\varphi_0$: 1 \\ $A$: linspace(0.001,8,120) \end{tabular} &  \\
  \hline \hline
GMSSMI & $V_0\left[\left(\frac{\varphi}{\varphi_0}\right)^2-\frac{2}
{3}\alpha\left(\frac{\varphi}{\varphi_0}\right)^6+\frac{\alpha}
{5}\left(\frac{\varphi}{\varphi_0}\right)^{10}\right]$ & \begin{tabular}{l} $\varphi_0$: 10.$\wedge$linspace(-2,0,3) \\ $\alpha$: [linspace(1e-2,2.5,120),10.$\wedge$linspace(0.5,1.5,3),10.$\wedge$linspace(-4,-2.5,6)] \end{tabular}  &  \\
  \hline \hline
GRIPI & $V_0\left[\left(\frac{\varphi}{\varphi_0}\right)^2-\frac{4}
{3}\alpha\left(\frac{\varphi}{\varphi_0}\right)^3+\frac{\alpha}
{2}\left(\frac{\varphi}{\varphi_0}\right)^{4}\right]$ & \begin{tabular}{l} $\varphi_0$: 10.$\wedge$linspace(-2,0,3) \\ $\alpha$: linspace(0.1,10,120) \end{tabular} & $[10^{-4},20]$ \\
  \hline \hline
BSUSYBI & $V_0\left(e^{\sqrt{6}\varphi} + e^{\sqrt{6} \gamma
  \varphi} \right)$ & $\gamma$: linspace(1e-5,2,200) & [-40,40] \\
  \hline \hline
TI & $V_0\left(1+\cos\frac{\varphi}{\mu}+\alpha\sin^2\frac{\varphi}{\mu}\right)$ & \begin{tabular}{l} $\mu$: 1 \\ $\alpha$: [linspace(0.01,3,80),10.$\wedge$linspace(-4,-2.5,6),10.$\wedge$linspace(1,2,4)] \end{tabular} &  \\
  \hline \hline
BEI & $V_0\exp_{1-\beta}\left(-\lambda \varphi\right)$ & \begin{tabular}{l} $\beta$: [linspace(-5,5,60),10.$\wedge$linspace(1,2,3),10.$\wedge$linspace(-4,-2,3), \\ -10.$\wedge$linspace(1,2,3),-10.$\wedge$linspace(-4,-2,3)] \\ $\lambda$: 1 \end{tabular} & [-100,100] \\
  \hline \hline
PSNI & $V_0\left[1+\alpha \ln \left(\cos\frac{\varphi}{f}\right)\right]$ & \begin{tabular}{l} $\alpha$: [linspace(0.1,10,60),10.$\wedge$linspace(-4,-1.5,12)] \\ $f$: 1 \end{tabular} & $[10^{-4},\pi/2-10^{-4}]$ \\
  \hline \hline
NCKI & $V_0\left[1+\alpha \ln \left(\varphi\right) + \beta
    \left(\varphi\right)^2\right]$ & \begin{tabular}{l} $\alpha$: 10.$\wedge$linspace(-7,0,16) \\ $\beta$: linspace(-10,10,80) \end{tabular} &  \\
  \hline \hline
CSI & $\frac{V_0}{\left( 1-\alpha\varphi \right)^2}$ & $\alpha$: [linspace(0.1,5,100),10.$\wedge$linspace(-4,-1.5,8),10.$\wedge$linspace(1,2,4)] & [-40,40] \\
  \hline \hline
OI & $V_0 \left(\frac{\varphi}{\varphi_0} \right)^{4}\left[
 \left(\ln \frac{\varphi}{\varphi_0} \right)^2- \alpha \right]$ & \begin{tabular}{l} $\varphi_0$: 1 \\ $\alpha$: [10.$\wedge$linspace(-7,-2,18),linspace(0.03,1,40)] \end{tabular} &  \\
  \hline \hline
CNCI & $V_0\left[ \left( 3+\alpha^2 \right) \coth^2
    \left(\frac{\alpha}{\sqrt{2}}\varphi\right)- 3 \right]$ & $\alpha$: [linspace(0.1,5,40),10.$\wedge$linspace(-7,-1.5,12),10.$\wedge$linspace(1,3,5)] &  \\
  \hline \hline
SBI & $V_0\left\lbrace 1 + \left[ -\alpha + \beta\ln
  \left( \varphi \right) \right] \left( \varphi
\right)^4 \right \rbrace$ & \begin{tabular}{l} $\alpha$: 10.$\wedge$linspace(-8,0,27) \\ $\beta$: 10.$\wedge$linspace(-8,0,27) \end{tabular} &  \\
  \hline \hline
SSBI & $V_0\left[1 + \alpha\left(\varphi\right)^2
  + \beta\left( \varphi \right)^4 \right]$ & \begin{tabular}{l} $\alpha$: [-10.$\wedge$linspace(-5,2,24),10.$\wedge$linspace(-5,2,24)] \\ $\beta$: [-10.$\wedge$linspace(-5,2,24),10.$\wedge$linspace(-5,2,24)] \end{tabular} & $[10^{-4},20]$ \\
  \hline \hline
IMI & $V_0\left(\varphi\right)^{-p}$ & $p$: linspace(0.5,10,40) &  \\
  \hline \hline
BI & $V_0 \left[1 -\left(\frac{\varphi}{\mu}\right)^{-p}\right]$  & \begin{tabular}{l} $p$: [linspace(1,10,37),10.$\wedge$linspace(-1,-0.33,3)] \\ $\mu$: [1e-4,1] \end{tabular} &  \\
  \hline \hline
RMI & $V_0\left[1-\frac{c}{2}\left(-\frac{1}{2} +\ln
\frac{\varphi }{\varphi_0}\right)\varphi ^2\right]$ & \begin{tabular}{l} $\varphi_0$: 1 \\ $c$: [-linspace(2,10,33),-10.$\wedge$linspace(-5,0,15),10.$\wedge$linspace(-5,0,14), \\ linspace(2,10,33)] \end{tabular} & $[10^{-4},10]$ \\
  \hline \hline
VHI & $V_0\left[1 +\left(\frac{\varphi}{\mu} \right)^{p} \right]$ & \begin{tabular}{l} $\mu$: 1 \\ $p$: linspace(0.1,12,80) \end{tabular} &  \\
  \hline \hline
DSI & $V_0\left[ 1+\left(\frac{\varphi}{\mu} \right)^{-p} \right]$ & \begin{tabular}{l} $\mu$: 1 \\ $p$: linspace(0.1,12,80) \end{tabular} &  \\
  \hline \hline
GMLFI & $V_0\left(\varphi\right)^p \left[1 + \alpha\left(
  \varphi \right)^q \right]$ & \begin{tabular}{l} $\alpha$: 10.$\wedge$linspace(-7,3,31) \\ $p$: linspace(0.5,12,24) \\ $q$: linspace(0.5,12,24) \end{tabular} & [-40,40] \\
  \hline \hline
LPI & $V_0\left(\frac{\varphi}{\varphi_0}\right)^{p}
  \left(\ln \frac{\varphi}{\varphi_0}\right)^q$ & \begin{tabular}{l} $\varphi_0$: 1 \\ $p$: [linspace(0.5,12,24),10.$\wedge$linspace(1.5,2,2)] \\ $q$: [linspace(0.5,12,24),10.$\wedge$linspace(1.5,2,2)] \end{tabular} &  \\
  \hline \hline
CNDI & $\frac{V_0}{ \left\lbrace 1 + \beta\cos\left[
    \alpha \left( \varphi-\varphi_0 \right) \right] \right \rbrace^2}$
 & \begin{tabular}{l} $\varphi_0$: 0 \\ $\alpha$: [linspace(0.1,1,30),10.$\wedge$linspace(-3,-1.5,4),10.$\wedge$linspace(0.5,2,4)] \\ $\beta$: [linspace(1,10,30),10.$\wedge$linspace(-2,-0.5,8),10.$\wedge$linspace(1.5,2,3), \\ -10.$\wedge$linspace(-2,1,7)] \end{tabular} &  \\  \hline
\caption{\footnotesize
Models from (Martin, Ringeval \& Vennin 2013) numerically tested.
Parameter values expressed in Matlab code: {\tt linspace(a,b,n)} means $n$ equispaced values between $a$ and $b$;
{\tt 10.$\wedge$linspace(a,b,n)} means $n$ log-equispaced values between $10^a$ and $10^b$.
Parameter $V_0$ and the reduced Planck mass $M_{\text{Pl}}$ always set to 1.
The range of studied field values is $\varphi \in [10^{-4},40]$ unless otherwise indicated.}
\end{longtable}}

The models, choice of parameter values and range of field values
 of Table \ref{t:llistamodels} have been subjected to the numerical
test described in subsection  
\ref{ss:algoritme} 
for the several determinations
of the spectral parameters $n_s,\alpha_s,\beta_s$. Let us sum up the
conclusions:

\medskip

 For the determination
of spectral parameters PLANCK2015 TT+lowP and  PLANCK2015 TT,TE,EE+lowP with running of the running
of Table 1,
the only models in Table \ref{t:llistamodels} which
are not disproved for any choice of parameter and field values
with confidence at least 92.8\% (the distance of the theoretical value of the runnig of the running to its mean observational data is larger than $1.8\sigma_{\beta}$) are:


\begin{enumerate}
\item Loop Infation (LI)  (Binetruy \& Dvali 1996; Halyo 1996)
\item   $R+R^{2p}$ Inflation  (RpI)  (Tsujikawa \& De Felice 2010; Nojiri \& Odintsov 2011).
\item K\"ahler Moduli Inflation II (KMIII)    (Colon \& Quevedo 2006; Lee \& Nam 2001)
\item Logamediate Inflation (LMI)  (Parson \& Barrow 1995; Barrow \& Nunes 2007).
\item Brane SUSY Breaking Inflation (BSUSYBI)  (Martin \& Ringeval 2004 ; Dudas et al. 2012).
\item Spontaneous Symmetry Breaking Inflation (SSBI)   (Albercht \& Brandenberger 1985; Hu and O'Connor 1986).
\item     Tip Inflation  (TI)  (Pajer 2008).
\item Generalised Mixed Large Field Inflation (GMLFI)  (Kinney \& Riotto 1998, 1999).
\item Constant $n_s$ D Inflation (CNDI)    (Hodges \& Blumenthal 1999).
\end{enumerate}

\noindent (albeit LI is disproved with 95.5\% confidence,
when one deals with the one-dimensional marginalized posterior distribution of the spectral index $n_s$). 


\section{Accuracy and reliability of the spectral parameter values} \label{s:reliability}

The computation of the spectral parameters from a single field, slow
roll theory with potential $V(\varphi)$, recalled in Sect. \ref{s:formules},
forecasts that the spectral index $n_s-1$ has order 1 on the inflation
parameters $\epsilon,\eta$, the running $\alpha_s$ has order 2 on $\epsilon, \eta$,
i.e. $\alpha_s=O((n_s-1)^2)$ because $\epsilon,\eta$ are small, 
and the running of the running $\beta_s$ has order 3, i.e. $\beta_s= O((n_s-1)^3)$.

The successive evaluations of the spectral parameters in  (Ade et al. 2013, 2015)
do not support this forecast:
\begin{enumerate}
\item The Planck 2013 determination without running of the running, finds $|n_s -1| \approx 4 \cdot 10^{-2}$, $\alpha_s \approx 
2 \cdot 10^{-2}$, the latter 2 standard deviations away from having a lower order.
\item The Planck 2015 determination without running of the running,  finds again $|n_s -1| \approx 4 \cdot 10^{-2}$, and now
$\alpha_s \approx 10^{-2}$, with standard deviation $\sigma \approx 10^{-2}$,
i.e. $\alpha_s$ is only  $1\sigma$ away from having a lower order.
\item But the Planck 2015 determination with running of the running, finds $|n_s -1| \approx 4 \cdot 10^{-2}$, $\alpha_s  \approx 10^{-2}$ with deviation $\sigma \approx 10^{-2}$,
and $\beta_s \approx 3 \cdot 10^{-2}$, with the running of the running  $2\sigma$ away 
from having a lower order.
\end{enumerate}

Therefore, the value of our simple analysis comparing model forecasts and 
experimental determinations depends on the reliability of these 
determinations of the values of the spectral parameters.

The methodology for this determination used in (Ade et al. 2013, 2015)
is based on model fitting through Bayesian statistical techniques and Markov-Monte Carlo 
(MCMC) optimization techniques. We have identified a flaw 
with the mathematical foundations of this procedure, which will lead to
incorrect values of the spectral parameters. 

This flaw comes from an inaccurate
interpretation of the meaning of the Taylor expansion of a function, and is independent 
of the statistical and optimization techniques used to fit the values, 
manifesting itself with similar consequences in computations with procedures 
ranging from least square optimization
to Bayesian likelihood maximization with MCMC methods. 
A very common manifestation of this flaw is the overestimation 
of the size of the highest order coefficient in the sought function.

We need to review the mathematical underpinnings of Taylor series expansions
and the method for the estimation
of the spectral parameters in order to explain the flaw.

The Taylor expansion of a function $g(\kappa)$ at a pivot scale $\kappa=0$
puts $g$ as a limit of a sequence of Taylor polynomials 
$S_d(g)=\sum_{i=0}^d \frac{g^{i)}(0)}{i!} \kappa^i$ with increasing degree $d$.
These polynomials are determined by the successive derivatives of $g$ at 0,
$g(0),g'(0),\dots, g^{d)}(0)$, and have the property that $S_d(g)$ is the
polynomial of a fixed maximal degree $d$ that best fits the values of the
function $g(\kappa)$ {\it for $\kappa$ in intervals $(-\delta,\delta)$ with 
$\delta << 1$}. That is, the fit of the polynomial $S_d$ to the function $g$
is optimal in a very small neighbourhood of $\kappa=0$, and becomes worse,
indeed irrelevant, for values of $\kappa$ outside this neighbourhood.  

When we try to approximate a function $g(\kappa)$ with a polynomial $p(\kappa)$
by minimizing a residue vector defined by the integrals over some family of test functions
$\{ \varphi_l \}$,
$$
R(p)=\left( \dots \int_0^{\kappa_f} \varphi_l(\kappa) p(\kappa) d\kappa -  
\int_0^{\kappa_f} \varphi_l(\kappa) g(\kappa) d\kappa \dots \right)
$$
the fit of the values of the polynomial $p(\kappa)$ to the function
$g(\kappa)$ is important {\it all over the integration interval $[0,\kappa_f]$}.
If the test functions in the family $\{ \varphi_l \}$ have its peaks and troughs
well spread over the integration interval, the fit of the values of
$p(\kappa)$ to the values of $g(\kappa)$ is of roughly equal importance
in all of the interval $[0,\kappa_f]$.

This is the mathematical reason why the polynomial of a fixed degree $d$ 
best approximating a
function $g(\kappa)$ through minimization of some residue defined over
an interval $[0,\kappa_f]$ will not be the Taylor polynomial of degree $d$ 
of $g$ at the pivot scale $\kappa=0$.

A concrete and frequent manifestation of this problem is that if one
tries to fit with a polynomial of a fixed degree $d$ the values of a function
$g(\kappa)$, over a fixed interval $[0,\kappa_f]$ where $g$ has order
of growth higher than $\kappa^d$, the residue--minimizing fit for
any reasonable evaluation of the residue will produce a polynomial
whose greatest degree coefficient is inflated in comparison
to the corresponding Taylor series coefficient of $g$,
in order for the leading term $a_d \kappa^d$ of the polynomial to approach a growth
of $g$ that actually has higher order.

We can illustrate this phenomenon with an elementary example: 
the approximation of the function $g(\kappa)=\kappa^4$ by taking
100 equispaced values of $g$ in the interval $[0,3]$, and finding the polynomial
$p_d$ of fixed degree $d$ that for the table of equispaced values $(\kappa_i,
g(\kappa_i))$ minimizes the residue 
$\sum_{i=1}^{100} \left( g(\kappa_i)-p_d(\kappa_i) \right)^2$. That is,
$p_d$ is the classical least square fit polynomial of degree $d$
for the tabulated values of $g$. The results are summarized in  Table 3.

\begin{table}
\begin{tabular}{c|ccccc} \label{t:interp}
degree $d$ of fit & $a_0$ & $a_1$ & $a_2$ & $a_3$ & $a_4$ \\
\hline
$\ge 4$ (exact values) & 0 & 0 & 0 & 0 & 1 \\
\hline
3  & -1.09 & 7.60 & -11.53 & 6 &  \\
\hline
2  & 6.77 & -24.64 & 15.47 & & \\
\hline
1  & -16.20 & 21.76 & & & \\
\hline
0  & 16.45 & & & &
\end{tabular}
\caption{ Regression polynomial and the overestimations for the
leading coefficients.}
\end{table}

As $g$ is a polynomial of degree 4, once the regression polynomial is allowed to reach this degree the least square polynomial $p_d$ is exactly $g$, and in
particular the coefficients of $p_d$ are exactly the coefficients of the 
Taylor expansion of $g$ at $\kappa=0$. But when a polynomial 
$p_d=a_0 +a_1 \kappa+ \dots + a_d \kappa^d$ of a degree $d<4$ is fit,
the least square fit results in a systematic bias: the leading coefficient 
$a_d$ is greatly overestimated compared to the coefficient of the same degree
in $g$, because the term $a_d \kappa^d$ follows values of $g$ with a growth
rate greater than $d$. This overestimation of the leading coefficient 
results in a cascade of alternating over- and underestimations for the
lower degree coefficients of the polynomial.

The same problem, further complicated by the fact that the sought function
is not polynomial and thus has infinitely many nonvanishing Taylor expansion coefficients,
happens with the 
 estimation of the spectral parameters of the power spectrum
through MCMC optimization (Ade et al. 2013, 2015):
one starts with the multipolar expansion of the CMB temperature
anisotropy over the last scattering sphere
\begin{eqnarray}
\frac{\delta T}{T}= \sum_{l \ge 1} \sum_{m=-l}^l a_{lm} Y_{lm},
\end{eqnarray}
which is determined by the angular power spectrum coefficients $\{ C_l \}_{l \ge 1}$
(see Kurki-Suonio (2005)), that can be computed, only taking into account temperature effects, for any given set of the cosmological  parameters $\vec \lambda$, as 
\begin{eqnarray} \label{eq:cl}
C_l(\vec\lambda) =   \frac{144 \pi^2}{25} \int k^2 dk\mathcal{P}_{\mathcal R}(k) | \Delta_{Tl}(k) |^2 \; ,
\end{eqnarray}
where the $\Delta_{Tl}(k)$ are suitable anisotropy transfer functions derived from the Boltzmann equation  (Seljak \& Zaldarriaga 1996).

We will recall now the definition, and perform a computation, for the spectral
parameters $n_s, \alpha_s, \beta_s$, to show how exactly the same 
bias that we have just discussed enters their determination. 
One initially selects a pivot scale $k_*$, puts the logarithm of the 
power spectrum $\ln {\mathcal P}_{\mathcal R}(k)$ as a function 
of a new variable $\kappa= \ln \left( \frac{k}{k_*} \right)$, and takes
the Taylor expansion at $\kappa=0$ 
\begin{equation} \label{eq:taylor}
\ln {\mathcal P}_{\mathcal R}(\kappa)= \ln A_s + (n_s-1) \kappa + \frac{1}{2} \alpha_s 
\kappa^2 + \frac{1}{6} \beta_s \kappa^3 + \dots
\end{equation}
The spectral coefficients are the succesive derivatives
$n_s-1=\frac{d}{d\kappa} \ln {\mathcal P}_{\mathcal R}(0)$,
$\alpha_s = \frac{d^2}{d\kappa^2} \ln {\mathcal P}_{\mathcal R}(0)$,
$\beta_s = \frac{d^3}{d\kappa^3} \ln {\mathcal P}_{\mathcal R}(0)$ \dots

In order to estimate de cosmological parameters $\vec\lambda$,
the computed $C_l(\vec\lambda)$ are compared to their observed values
$\widehat{C_l}$, and the values attributed to the cosmological parameters, the best fit values, are those that minimize an error function, 
the so-called likelihood function $\mathcal L$, defined by $\chi^2\equiv -2\ln {\mathcal L}$, where the chi-square is a suitable quadratic ``distance'' between $C_l(\vec\lambda)$ and
$\widehat{C_l}$.
The classical chi-square   (Press 1992), used in  (Dodelson, Kinney \& Kolb 1997), is 
\begin{eqnarray} \label{eq:chi2}
\chi^2(\vec\lambda)= \sum_l \frac{(\widehat{C_l}-C_l(\vec\lambda))^2}{\sigma_l^2} \, ,
\end{eqnarray}
and is basically the residue of a least square fit, measured in units of deviation $\sigma$.
There are variants for the definition of the chi-square function, such as
that in  (Verde et al. 2003) ,
\begin{eqnarray} \label{eq:verde}
 \chi^2(\vec\lambda)= \sum_l (2l+1)\left[\ln\left(\frac{C_l(\vec\lambda)}{\widehat{C_l}}\right)+\frac{\widehat{C_l}}{C_l(\vec\lambda)}-1\right].
\end{eqnarray}
 but for close fits (i.e. $\frac{\widehat{C_l}-C_l}{\widehat{C_l}}$
small) the variants have similar values and lead to a similar optimization result.

In all cases, due to the complicated form of $C_l(\vec\lambda)$, the minimizatition 
is determined by any suitable numerical method, such as MCMC  (Christensen \& Meyer 2000).

We can perform now, as a numerical experiment, an instance of the central part of this computation, in order to show how the
bias is introduced in it. Let us assume a single field, slow roll Universe
in which the values of all cosmological
parameters in $\vec{\lambda}$ are known, and in particular has a power
spectrum of the form
\begin{eqnarray} \label{eq:pllarg}
\ln {\mathcal P}_{\mathcal R}(\kappa)= \ln A_s + (n_s-1) \kappa + \frac{1}{2} \alpha_s 
\kappa^2 + \frac{1}{6} \beta_s \kappa^3 \\ +  \frac{1}{24} \gamma_s \kappa^4 + \frac{1}{120} \delta_s \kappa^5 + \frac{1}{720} \zeta_s \kappa^6 + \frac{1}{5040} \theta_s \kappa^7
\end{eqnarray}
with parameter values $n_s=0.96, \alpha_s=6 \cdot 10^{-3}, \beta_s=2 \cdot 10^{-4},
\gamma_s=5 \cdot 10^{-5}, \delta_s=8 \cdot 10^{-6}, \zeta_s=10^{-7}, 
\theta_s=10^{-8}$. The parameter values have
decreasing order in agreement with the theoretical forecast.

With this power spectrum, we perform the computation of the coefficients $C_l$
for $l$ ranging from 10 to 1400 with step 10 according to the formula \eqref{eq:cl}.
The selected transfer function $\Delta_{Tl}(k)$ is that of the CMBSimple procedure
of  (Baumann 2011). This uses the two fluid approximation of Seljak, and being
a stripped-down version of standard codes such as CAMB or CMBFast it allows
a clear exhibition of the computation bias we wish to show. Figure $1$
shows the resulting values for the coefficients $C_l$.

\begin{figure}[h]\label{f:cld}
\begin{center}
\includegraphics[scale=0.5]{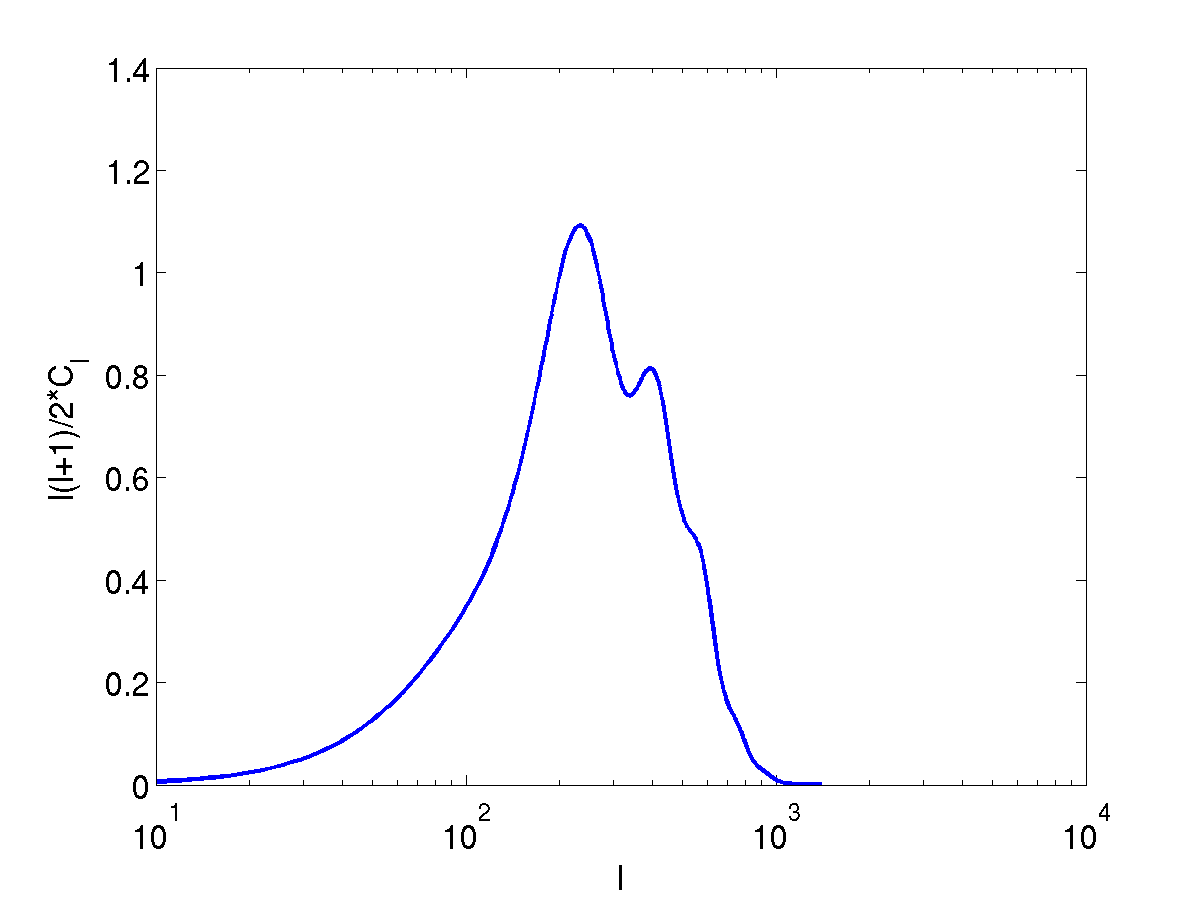}
\end{center}
\caption{ Coefficients $C_l$ using the CMBSimple.}
\end{figure}

Let us assume now that these values of the coefficients $C_l$ are the observed
values, and that we know exactly every cosmological parameter other than the 
spectral coefficients of the power spectrum ${\mathcal P}_{\mathcal R}(\kappa)$.
We will perform the inverse computation, in which the spectral coefficients
$n_s, \alpha_s$ and possibly $\beta_s, \gamma_s, \delta_s$ are given the values that minimize either the
residue \eqref{eq:chi2}, or that in \eqref{eq:verde}. As the value of every
other cosmological parameter is known, the optimization problem is now simple 
enough so that several standard algorithms
such as MCMC, the Levenberg-Marquardt method for nonlinear regression
or the simplex search method for minimization of the residue function (Press 1992)
will converge to the sought best parameter fit.

Table \ref{t:coefs} shows in its columns the results of 3 computations, finding the
spectral parameters in a power spectrum that has terms up to $\alpha_s$, 
resp. up to $\beta_s$, up to $\delta_s$, by computing 
the coefficients $C_l$ for $l$ ranging from 10 to 1400 with step 10 according to
Eq. \eqref{eq:cl} for each choice of 
spectral parameter value and minimizing the residue in formula \eqref{eq:verde}
in the comparison with the "observed" values. Minimizing 
the residue \eqref{eq:chi2} yields roughly the same values for the spectral
parameters.

\begin{table}
\begin{tabular}{c|c|c|c|c} \label{t:coefs}
$\downarrow$ {\small param.}  {\small computation} $\rightarrow$ & 
{\small correct value} & {\small fit up to $\alpha_s$} & {\small up to $\beta_s$} & 
{\small up to $\delta_s$} \\
\hline
$n_s$ & 0.96 & 0.9374 & 0.9708 & 0.9653 \\
\hline
$\alpha_s$  & $6\cdot 10^{-3}$ & $1.14\cdot 10^{-2}$ & $1.34\cdot 10^{-3}$ & 
$2.64\cdot 10^{-3}$ \\
\hline
$\beta_s$  & $2\cdot 10^{-4}$ &  & $1.12\cdot 10^{-3}$ & $1.6\cdot 10^{-3}$ \\
\hline
$\gamma_s$  & $5\cdot 10^{-5}$ &  &  & $-1.95\cdot 10^{-4}$ \\
\hline
$\delta_s$  & $8\cdot 10^{-6}$ & & & $3.37\cdot 10^{-5}$
\end{tabular}
\caption{}
\end{table}

\medskip

The results in Table \ref{t:coefs} are strikingly similar to those in
Table $3$.
Our recurring topic in this section of the paper perfectly interprets the
results of these spectral parameter determinations:
\begin{itemize}
\item In each computation, the highest order parameter that has been used for
fitting the $C_l$ coefficients has been overestimated: by a factor of 1.9
in the computation up to $\alpha_s$, by a factor of 5.5 in the computation 
up to $\beta_s$, by a factor of 4.2 in the computation up to $\delta_s$.
\item The second to highest order parameter has been underestimated 
in each computation, even changing sign for $\gamma_s$ in the last one,
as a correction to the overestimation of the highest order term.
\item As the computation of spectral parameters assumes a polynomial function
$\ln {\mathcal P}_{\mathcal R}(\kappa)$ with coefficients up to higher order,
the determination of the lowest order coefficients $n_s,\alpha_s$, becomes more accurate. This is indeed a trend, but the greater complexity and 
propagation of measurement errors of the residue
minimization as the number of parameters grows do not allow in practice
a scheme in which $\ln {\mathcal P}_{\mathcal R}(\kappa)$ is determined
with coefficients up to a high order, so as to obtain reliable values 
for the low order coefficients. 
\end{itemize}

These conclusions hold for the best fit solution to the optimization 
problem of minimizing the residue. 
Therefore they hold for any procedure that may be employed
to find the best fit parameters (MCMC, \dots ).

\section{Conclusions}

Single field slow rolling inflaton
models do not fit well the computation of the spectral parameters of the 
temperature power spectrum of the CMB radiation by the Planck collaboration
(Ade et al. 2013, 2015). 

Of the 49 models examined in the study (Martin, Ringeval \& Vennin 2013),
only the following 8 can, with any choice of parameters for the model, 
yield values for the spectral parameters $n_s, \alpha_s, \beta_s$ 
at a distance of less than $1.8 \sigma$ from the values given by 
(Ade et al. 2013, 2015). The models that do not lie outside this
92.8\% confidence interval are RpI, KMIII, LMI, BSUSYBI, SSBI, TI, 
GMLFI, CNDI (model details and references in \ref{ss:numresults}).

The ultimate reason for this poor fit is that the reported values 
for the parameters $n_s-1, \alpha_s, \beta_s$ does not follow the hierarchy 
of sizes forecast by the slow roll theory, according to which these parameters
ought to have decreasing order of magnitude. The running of the running
$\beta_s$ reported by PLANCK2015 has a size $O(10^{-2})$ which is so large
 that most of the tested inflaton models cannot furnish values
within 1.8 standard deviations of the expected value.

This discrepancy between the values for the spectral parameters determined
by the Planck collaboration and the expected slow roll size hierarchy
for them can arguably be due to lack of accuracy in the determination
of these values, specially for the higher order parameters. Such lack
of accuracy would mean that the models that do not match the used determinations
of $\alpha_s, \beta_s$ are not disproved by this mismatch.

In this work we identify a bias in the method that has been used for 
the computation of the
values of the spectral parameters, which results in a systematic overestimation
of the highest order one. The bias is a migration of, and very close to,
a classical bias of regression (i.e., minimization of residue) fitting:
if one tries to fit a polynomial
of too low degree to values of a function that actually grows faster,
no matter what the procedure for finding the better fit is,
it will result in a polynomial with an exaggerated value for the magnitude of the
leading term.

While vastly more powerful than their predecessors, the probabilistic 
(Bayesian, MCMC, \dots) methods currently used to
fit the value of spectral parameters following a Gaussian distribution
in a model ultimately choose the fit that minimizes the residue, 
 and inherit this bias. 

Removal of this bias
will be necessary to judge if single field, slow roll inflation models 
match the increasingly refined measuremnents of the CMB, or if
sophistications
such as scalar electrodynamics and $SU(5)$ RG-improved potentials 
(Elizalde, Odintsov, Pozdeeva \& Vernov 2014),
warm inflation  (Berera 1995; Bastero-Gil 2014), multiple fields, a breakdown in the slow roll regime (Easther \& Peiris 2006; Wan et al. 2014), or a
completely different paradigm such as the Matter Bounce Scenario can be contemplated (Cai, Brandenberger \& Zhang 2009, 2011; Cai 2014).

The authors have little hope that spectral parameters such as the running $\alpha_s$,
the running of the running $\beta_s$, and the higher order coefficients of
the power spectrum ${\mathcal P}_{\mathcal R}$ can be computed reliably 
from experimental observations. These invariants are derivatives of increasingly
high order of the function's logarithm at a pivot value $k=k_*$, and 
derivatives are notoriously hard to compute in the presence of noise
(i.e., experimental errors and slow roll approximations)
in the data. Indeed, schemes to compute these spectral parameters up to a high
order by residue minimization so as to get a reliable determination of
the lowest order ones fail because of the propagation of the noise 
throughout the computation.

An approach that seems more promising is to avoid these spectral parameters:
the Mukhanov-Sasaki equations
\begin{eqnarray}
v_k''+ \left( k^2-\frac{z''}{z} \right) v_k=0, \qquad z=a \frac{\varphi'}{\mathcal H}
\end{eqnarray}
together with the conservation and Friedmann equations
\begin{eqnarray}
\ddot{\varphi} + 3H \varphi + V_{\varphi}=0, \\
H=\frac{1}{\sqrt{3}} \sqrt{ \frac{\dot \varphi^2}{2}+ V(\varphi) },
\end{eqnarray}
determine the power spectrum ${\mathcal P}_{\mathcal R}$ once the parameters
$(\varphi_0,\dot{\varphi}_0)$ on which the background depends as initial
conditions for the conservation equation, and those
required by the potential $V(\varphi)$, have been set.
The coefficients $C_l$ can then be determined from ${\mathcal P}_{\mathcal R}$,
and one can look for the values of the background and potential parameters
yielding a power spectrum ${\mathcal P}_{\mathcal R}$
that minimizes the residue \eqref{eq:verde}, or \eqref{eq:chi2}.
This is a computationally intensive task, so an efficient residue minimization
scheme such as MCMC seems appropiate. The fit of particular models
to the observations of $C_l$ can be judged, e.g. by means of Bayesian
techniques, through these background and potential parameters.
The authors believe that this approach merits further investigation.

\medskip

This investigation has been supported in part by MINECO (Spain), projects MTM2011-27739-C04-01, and MTM2012-38122-C03-01.

\end{document}